\documentclass[twocolumn,showpacs,preprintnumbers,amsmath,amssymb]{revtex4}
\usepackage{graphicx}% Include figure files
\usepackage{epsfig}
\usepackage{dcolumn}% Align table columns on decimal point
\usepackage{bm}% bold math

\def\aA{$\alpha$-nucleus }
\def\AA{nucleus-nucleus }
\def\ac12{$\alpha+^{12}$C}
\def\sigR{$\sigma_{\rm R}$}
\def\sigI{$\sigma_{\rm I}$}
\def\rmsr{$<r^2>^{1/2}$}
\def\rmsp{$<r^2>^{1/2}_{p}$}
\def\rmsn{$<r^2>^{1/2}_{n}$}
\def\rmsexp{$<r^2>^{1/2}_{\rm exp}$}
\def\rmscalc{$<r^2>^{1/2}_{\rm calc}$}
\begin{document}
%\preprint{Draft for Phys. Rev. C}
% \preprint{APS/123-QED}
\title{Microscopic calculation of the interaction cross section for stable and
unstable nuclei based on the nonrelativistic nucleon-nucleon $t$-matrix}
\author{Dao T. Khoa}\email{khoa@vaec.gov.vn}
\author{Hoang Sy Than}
\affiliation{Institute for Nuclear Science {\rm \&} Technique,
VAEC, P.O. Box 5T-160, Nghia Do, Hanoi, Vietnam.}
\author{Tran Hoai Nam}
\affiliation{Department of Physics, Hanoi University of Natural
Sciences, 334 Nguyen Trai Str., Thanh Xuan, Hanoi, Vietnam.}
\author{Marcella Grasso and Nguyen Van Giai}
\affiliation{Institut de Physique Nucl\'eaire, IN2P3-CNRS, 91406
Orsay Cedex, France.}
\date{Accepted for publication in Physical Review C}% It is always \today, today,
\begin{abstract}
Fully quantal calculations of the total reaction cross sections \sigR\ and
interaction cross sections \sigI, induced by stable and unstable He, Li, C and O
isotopes on $^{12}$C target at $E_{\rm lab}\approx 0.8$ and 1 GeV/nucleon have
been performed, for the first time, in the distorted wave impulse approximation
(DWIA) using the microscopic \emph{complex} optical potential and inelastic form
factors given by the folding model. Realistic nuclear densities for the
projectiles and $^{12}$C target as well as the complex $t$-matrix
parameterization of free nucleon-nucleon interaction by Franey and Love were
used as inputs of the folding calculation. Our \emph{parameter-free} folding +
DWIA approach has been shown to give a very good account (within 1--2\%) of the
experimental \sigI\ measured at these energies for the stable, strongly bound
isotopes. With the antisymmetrization of the dinuclear system properly taken
into account, this microscopic approach is shown to be more accurate than the
simple optical limit of Glauber model that was widely used to infer the nuclear
radii from the measured \sigI. Therefore, the results obtained for the nuclear
radii of neutron-rich isotopes under study can be of interest for further
nuclear structure studies.
\end{abstract}
\maketitle

\section{Introduction}
Since the 80s of the last century, the radioactive ion beams have been used
intensively to measure the total reaction cross sections and interaction cross
sections induced by unstable nuclei on stable targets (see a recent review in
Ref.~\cite{Oz01}) which serve as an important data bank for the determination of
nuclear sizes. The discovery of exotic structures of unstable nuclei, such as
neutron halos or neutron skins, are among the most fascinating results of this
study.

The theoretical tool used dominantly by now to analyze the interaction cross
sections measured at energies of several hundred MeV/nucleon is the Glauber
model \cite{Gl70,Og92} which is based on the eikonal approximation. This
approach provides a simple connection between the ground state densities of the
two colliding nuclei and the total reaction cross section of the \AA system and
has been used, in particular, to deduce the nuclear density parameters for the
neutron-rich halo nuclei \cite{Ta96}.

In general, the total reaction cross section \sigR, which measures the loss of
flux from the elastic channel, must be calculated from the transmission
coefficient $T_l$ as
\begin {equation}
\sigma_{\rm R}=\frac{\pi}{k^2}\sum_l(2l+1)T_l,
\label{e1}
\end{equation}
where $k$ is the relative momentum (or wave number). The summation is carried
over all partial waves $l$ with $T_l$ determined from the elastic $S$-matrix as
\begin {equation}
T_l=1-|S_l|^2.
\label{e2}
\end{equation}
In the standard optical model (OM), the quantal $S$-matrix elements $S_l$ are
obtained from the solution of the Schr\"odinger equation for elastic \AA
scattering using a \emph{complex} optical potential. At low energies, the
eikonal approximation is less accurate and, instead of Glauber model, the OM
should be used to calculate \sigR for a reliable comparison with the data. At
energies approaching 1 GeV/nucleon region, there are very few elastic scattering
data available and the choice of a realistic optical potential becomes
technically difficult, especially for unstable nuclei. Perhaps, that is the
reason why different versions of Glauber model are widely used to calculate
\sigR\ at high energies. Depending on the structure model for the nuclear wave
functions used in the calculation, those Glauber model calculations can be
divided into two groups: the calculations using a simple optical limit of
Glauber model (see Ref.~\cite{Oz01} and references therein) and the more
advanced approaches where the few-body correlation and/or breakup of a
loosely-bound projectile into a core and valence (halo) nucleons are treated
explicitly \cite{Og92,Al96,To98}.

In the present work, we explore the applicability of the standard OM to
calculate the total reaction cross section (\ref{e1}) induced by stable and
unstable beams at high energies using the microscopic optical potential
predicted by the folding model. The basic inputs of a folding calculation are
the densities of the two colliding nuclei and the effective nucleon-nucleon (NN)
interaction \cite{Sa79}. At low energies, a realistic density dependent NN
interaction \cite{Kh97} based on the M3Y interaction \cite{Be77} has been
successfully used to calculate the \aA and \AA optical potential \cite{Kh00}.
This interaction fails, however, to predict the shape of the \aA optical
potential as the bombarding energy increases to about 340 MeV/nucleon
\cite{Kh02}. On the other hand, at incident energies approaching a few hundred
MeV/nucleon the $t$-matrix parameterization of free NN interaction was often
used in the folding analysis of proton-nucleus scattering \cite{Lo81,Fr85}. The
use of the $t$-matrix interaction corresponds to the so-called \emph{impulse
approximation} (IA), where the medium modifications of the NN interaction are
neglected \cite{Sa83}.

In the present folding calculation we adopt a local representation of the free
NN $t$-matrix developed by Franey and Love \cite{Fr85} based on the experimental
NN phase shifts. The folded optical potentials and inelastic form factors are
used further in the distorted wave impulse approximation (DWIA) to calculate
\sigR\ and interaction cross section \sigI, induced by stable and unstable He,
Li, C, and O isotopes on $^{12}$C target at bombarding energies around 0.8 and 1
GeV/nucleon. Since relativistic effects are significant at high energies, the
relativistic kinematics are taken into account properly in both the folding and
DWIA calculations. To clarify the adequacy and possible limitation of the
present folding model, we also discuss the main approximations made in our
approach and compare them with those usually assumed in the Glauber model.

Given the realistic nuclear densities and validity of IA, the folding approach
presented below in Sec.~\ref{sec2} is actually parameter-free and it is
necessary to test first the reliability of the model by studying the known
stable nuclei before going to study unstable nuclei. Such a procedure is
discussed briefly in Sec.~\ref{sec3}. Then, \sigI\ measured for the neutron-rich
He, Li, C, and O isotopes are compared with the results of calculation and the
sensitivity of nuclear radii to the calculated \sigI\ is discussed. The
discrepancy between $\sigma^{\rm calc}_{\rm I}$ and $\sigma^{\rm exp}_{\rm I}$
found for some light halo nuclei is discussed in details to indicate possible
effects caused by the dynamic few-body correlation. Conclusions are drawn in
Sec.~\ref{sec4}.

\section{Folding model for the complex \AA optical potential}
 \label{sec2}
The details of the latest double-folding formalism are given in Ref.~\cite{Kh00}
and we only recall briefly its main features. In general, the projectile-target
interaction potential can be evaluated as an energy dependent Hartree-Fock-type
potential of the dinuclear system
\begin {equation}
 U=\sum_{i\in a,j\in A}[<ij|v_{\rm D}|ij> +<ij|v_{\rm EX}|ji>]
 =V_{\rm D}+V_{\rm EX},
 \label{e5}
\end{equation}
where the nuclear interaction $V$ is a sum of effective NN interactions $v_{ij}$
between nucleon $i$ in the projectile $a$ and nucleon $j$ in the target $A$. The
antisymmetrization of the dinuclear system is done by taking into account the
single-nucleon knock-on exchanges.

The direct part of the potential is local (provided that the NN interaction
itself is local), and can be written in terms of the one-body densities,
\begin{eqnarray}
 V_{\rm D}(E,\bm{R})=\int\rho_a(\bm{r}_a)\rho_A(\bm{r}_A)
 v_{\rm D}(E,\rho,s)d^3r_a d^3r_A,\nonumber\\
 \ \ {\rm where}\ \bm{s}=\bm{r}_A-\bm{r}_a+\bm{R}.
 \label{e6}
\end{eqnarray}
The exchange part is, in general, nonlocal. However, an accurate local
approximation can be obtained by treating the relative motion locally as a plane
wave \cite{Si79}
\begin{eqnarray}
 V_{\rm EX}(E,\bm{R})=\int\rho_a(\bm{r}_a,\bm{r}_a+\bm{s})
 \rho_A(\bm{r}_A,\bm{r}_A-\bm{s}) \nonumber\\
 \times\ v_{\rm EX}(E,\rho,s)\exp\left(\frac{i\bm{K}(E,\bm{R})\cdot \bm{s}}{M}\right)
 d^3r_ad^3r_A.
 \label{e7}
\end{eqnarray}
Here $\rho_a(\bm{r}_a)\equiv\rho_a(\bm{r}_a,\bm{r}_a)$ and
$\rho_a(\bm{r}_a,\bm{r}_a+\bm{s})$ is the diagonal and nondiagonal parts of the
one-body density matrix for the projectile, and similarly for the target.
$\bm{K}(E,\bm{R})$ is the local momentum of relative motion determined as
\begin {equation}
 K^2(E,\bm{R})=\frac{2\mu}{\hbar^2}[E_{\rm c.m.}-{\rm Re}\ U(E,\bm{R})
 -V_{\rm C}(\bm{R})],
\label{e8}
\end{equation}
$\mu$ is the reduced mass, $M=aA/(a+A)$ with $a$ and $A$ the mass numbers of the
projectile and target, respectively. Here, $U(E,\bm{R})=V_{\rm
D}(E,\bm{R})+V_{\rm EX}(E,\bm{R})$ and $V_{\rm C}(\bm{R})$ are the total
nuclear and Coulomb potentials, respectively. More details on the calculation of
the direct and exchange potentials (\ref{e6})-(\ref{e7}) can be found in
Refs.~\cite{Kh00,Kh01}. The folding inputs for mass numbers and incident
energies were taken as given by the relativistically corrected kinematics
\cite{Fa84}.

To calculate consistently both the optical potential and inelastic form factor
one needs to take into account explicitly the multipole decomposition of the
nuclear density that enters the folding calculation \cite{Kh00}
\begin {equation}
 \rho_{JM\to J'M'}(\bm{r})=\sum_{\lambda\mu}
 <JM\lambda\mu|J'M'>C_\lambda\rho_\lambda(r)
 [i^{\lambda}Y_{\lambda\mu}(\hat{\bm{r}})]^*,
\label{e10}
\end{equation}
where $JM$ and $J'M'$ are the nuclear spin and its projection in the initial and
final states, respectively, and $\rho_\lambda(r)$ is the nuclear transition
density for the corresponding $2^{\lambda}$-pole excitation. In the present
work, we adopt the collective-model Bohr-Mottelson prescription \cite{Bo75} to
construct the nuclear transition density for a given excitation in the $^{12}$C
target as
\begin {equation}
 \rho_\lambda(r)=-\delta_\lambda\frac{d\rho_0(r)}{dr}.
 \label{e11}
\end{equation}
Here $\rho_0(r)$ is the total ground state (g.s.) density and $\delta_\lambda$
is the deformation length of the $2^{\lambda}$-pole excitation in the $^{12}$C
target.

\subsection*{Impulse approximation and the $t$-matrix interaction}
If the total spin and isospin are zero for one of the two colliding nuclei
($^{12}$C in our case) only the spin- and isospin-independent components of the
central NN forces are necessary for the folding calculation. We discuss now the
choice of $v_{\rm D(EX)}(E,\rho,s)$ for the two bombarding energies of 0.8 and 1
GeV/nucleon. At these high energies, one can adopt the IA which reduces the
effective NN interaction approximately to that between the two nucleons in
vacuum \cite{Sa83}. Consequently, the microscopic optical potential and
inelastic form factors can be obtained by folding the g.s. and transition
densities of the two colliding nuclei with an appropriate $t$-matrix
parameterization of the free NN interaction.

In the present work, we have chosen the nonrelativistic $t$-matrix interaction
which was developed by Franey and Love \cite{Fr85} based on experimental NN
phase shifts at bombarding energies of 0.8 and 1 GeV. The spin- and
isospin-independent direct ($v_{\rm D}$) and exchange ($v_{\rm EX}$) parts of
the central NN interaction are then determined from the singlet- and
triplet-even (SE,TE) and odd (SO,TO) components of the local $t$-matrix
interaction (see Table I of Ref.~\cite{Fr85}) as
\begin {equation}
 v_{\rm D(EX)}(s)=\frac{k_ak_A}{16}[3t_{\rm TE}(s)+3t_{\rm SE}(s)\pm
 9t_{\rm TO}(s)\pm 3t_{\rm SO}(s)].
 \label{v1}
\end {equation}
Here $k_a$ and $k_A$ are the energy-dependent kinematic modification factors of
the $t$-matrix transformation \cite{Ke59} from the NN frame to the N$a$ and N$A$
frames, respectively. $k_a$ and $k_A$ were evaluated using Eq.~(19) of
Ref.~\cite{Lo81}. The explicit, complex strength of the \emph{finite-range}
central $t$-matrix interaction (\ref{v1}) is given in terms of four Yukawas
\cite{Fr85}. Since the medium modifications of the NN interaction are neglected
in the IA \cite{Sa83}, the $t$-matrix interaction (\ref{v1}) does not depend on
the nuclear density.

\subsection*{Main steps in the calculation of $\bm{\sigma_{\rm\bf I}}$}
With properly chosen g.s. densities for the two colliding nuclei, the elastic
scattering cross section and \sigR\ are obtained straightforwardly in the OM
calculation using the microscopic optical potential (\ref{e6})-(\ref{e8}). We
recall that the interaction cross section \sigI\ is actually the sum of all
particle removal cross sections from the projectile \cite{Oz01} and accounts,
therefore, for all processes when the neutron and/or proton number in the
\emph{projectile} is changed. As a result, \sigI\ must be smaller than the total
reaction cross section \sigR\ which includes also the cross section of inelastic
scattering to excited states in both the target and projectile as well as cross
section of nucleon removal from the target. At energies of several hundred
MeV/nucleon, the difference between \sigR\ and \sigI\ was found to be a few
percent \cite{Og92,Kh95b,Ja78} and was usually neglected to allow a direct
comparison of the calculated $\sigma_{\rm R}$ with the measured \sigI. Since the
experimental uncertainty in the measured \sigI\ is very small at the considered
energies (around 1\% for stable projectiles like $^{4}$He, $^{12}$C, and
$^{16}$O \cite{Oz01}) neglecting the difference between \sigR\ and \sigI\ might
be too rough an approximation in comparing the calculated \sigR\ with the
measured \sigI\ and testing nuclear radius at the accuracy level of $\pm 0.05$
fm or less \cite{Oz01,Oza01}. In the present work, we try to estimate \sigI\ as
accurately as possible by subtracting from the calculated \sigR\ the total cross
section of the main inelastic scattering channels. Namely, we have calculated in
DWIA, using the complex folded optical potential and inelastic form factors, the
integrated cross sections $\sigma_{2^+}$ and $\sigma_{3^-}$ of inelastic
scattering to the first excited 2$^+$ and 3$^-$ states of $^{12}$C target at
4.44 and 9.64 MeV, respectively. These states are known to have the largest
cross sections in the inelastic proton and heavy ion scattering on $^{12}$C at
different energies. The deformation lengths used to construct transition
densities (\ref{e11}) for the folding calculation were chosen so that the
electric transition rates measured for these states are reproduced with the
proton transition density as
\begin {equation}
B(E\lambda\uparrow)=e^2\Big|\int_0^\infty \rho_{\lambda}^p(r)
r^{\lambda+2}dr\Big|^2.
\label{s1}
\end {equation}
Using a realistic Fermi distribution for the g.s. density of $^{12}$C (see next
Section) to generate the transition densities, we obtain $\delta_2\approx 1.54$
fm and $\delta_3\approx 2.11$ fm which reproduce the experimental transition
rates $B(E2\uparrow)\approx 41\ e^2$fm$^4$ \cite{Ra01} and $B(E3\uparrow)\approx
750\ e^2$fm$^6$ \cite{Sp89}, respectively, via Eq.~(\ref{s1}). Since inelastic
scattering to excited states of the unstable projectile is suppressed by a much
faster breakup process, \sigI\ can be approximately obtained as
\begin{eqnarray}
 \sigma_{\rm I} & = & \sigma_{\rm R}-\sigma_{\rm Inel} \nonumber \\
 & \approx & \sigma_{\rm R}-\sigma_{2^+}-\sigma_{3^-}.
\label{s2}
\end{eqnarray}
All the OM and DWIA calculations were made using the code ECIS97 \cite{Ra97}
with the relativistic kinematics properly taken into account. At the energies
around 1 GeV/nucleon the summation (\ref{e1}) is usually carried over up to 800
- 1000 partial waves to reach the full convergence of the $S$-matrix series for
the considered \AA systems.

\subsection*{Adequacy and limitation of the folding approach}
Since the measured \sigI\ have been analyzed extensively by different versions
of Glauber model and its optical limit (OL) is sometimes referred to as the
folding model \cite{To98,Es99}, we find it necessary to highlight the
distinctive features of the present folding approach in comparison with the OL
of Glauber model before going to discuss the results of calculation.

\begin{figure*}[htb]\vspace*{-2.5cm}
\begin{minipage}[t]{8.5cm}
\hspace*{-1cm} %\vspace*{-3cm}
 \mbox{\epsfig{file=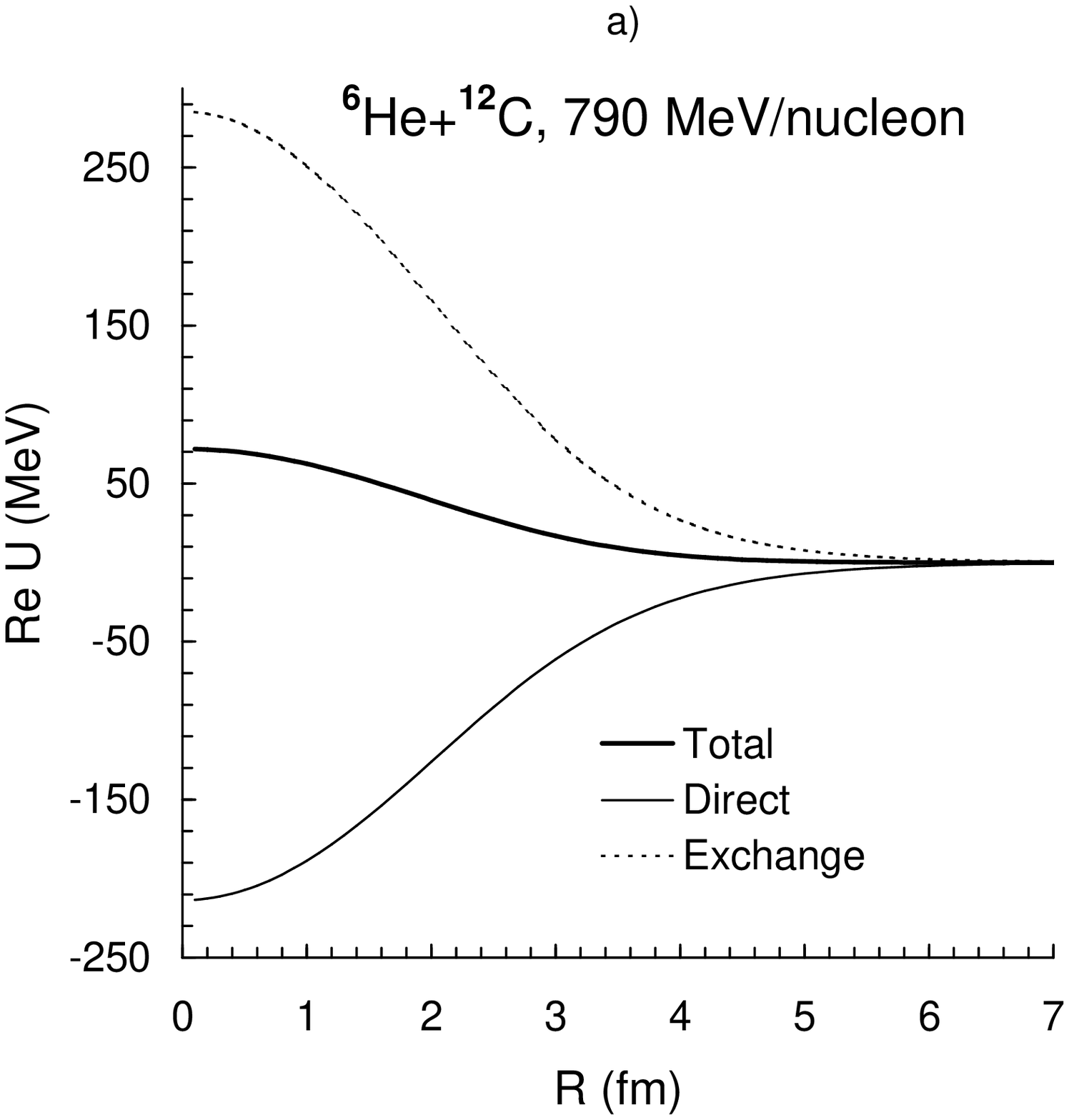,height=15cm}}
\end{minipage}
\hspace{\fill}
\begin{minipage}[t]{8.5cm}
\hspace*{-2cm} %\vspace*{-1cm}
 \mbox{\epsfig{file=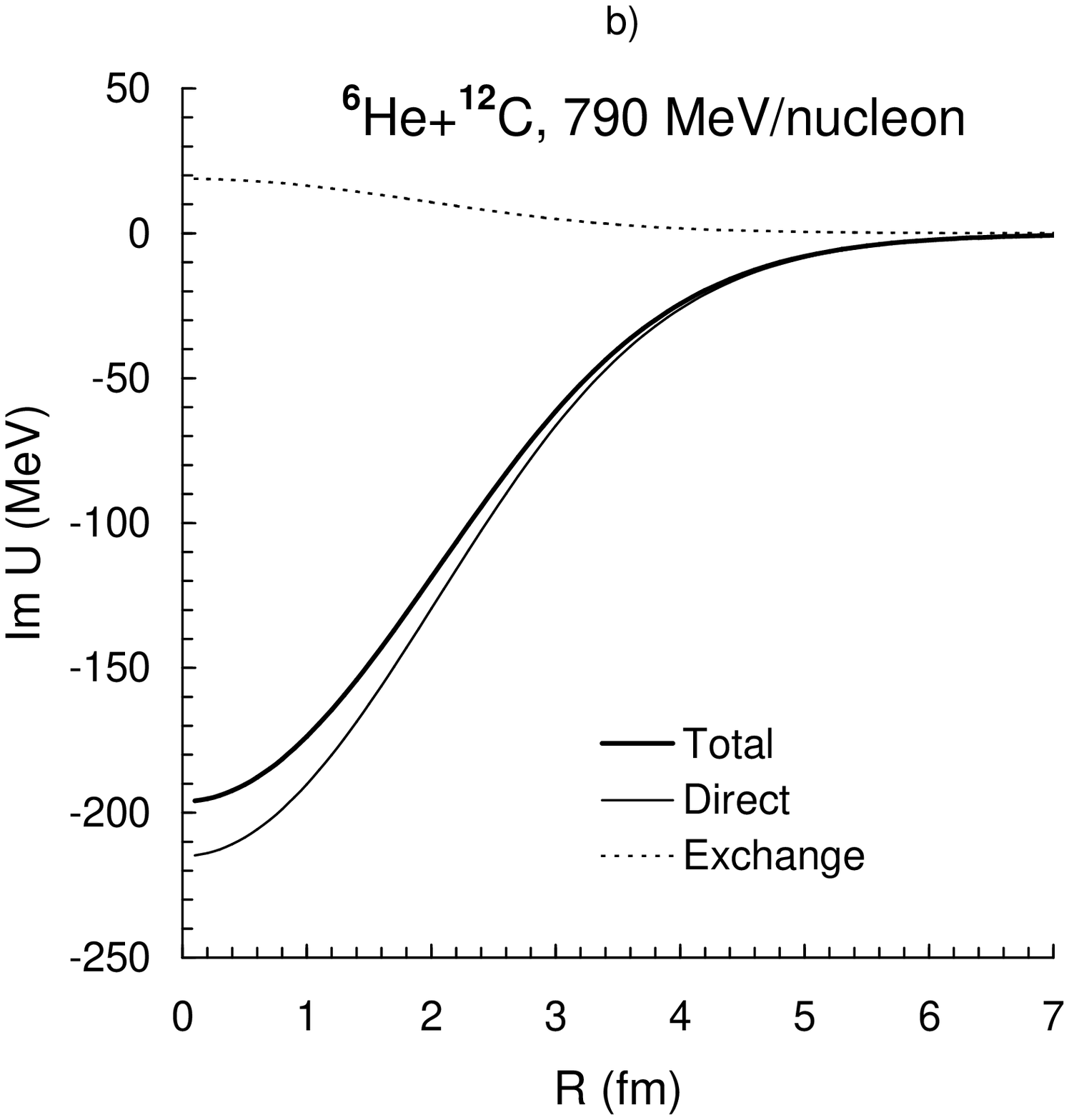,height=15cm}}
\end{minipage}\vspace*{-3cm}
\caption{Radial shape of the \emph{direct} $V_{\rm D}$ and \emph{exchange}
$V_{\rm EX}$ parts of the \emph{total} optical potential $U$ for
$^{6}$He+$^{12}$C system at 790 MeV/nucleon. The real and imaginary parts of
$U$ are shown in panels \emph{a} and \emph{b}, respectively.} \label{f0}
\end{figure*}

On the level of the \AA optical potential (OP), the present double-folding
approach evaluates OP using fully \emph{finite-range} NN interaction and taking
into account the exchange effects accurately via the Fock term in
Eq.~(\ref{e5}). Therefore, individual nucleons are allowed to scatter after the
collision into unoccupied single-particle states only. Sometimes, one discusses
these effects as the exchange NN correlation. An appropriate treatment of the
exchange NN correlation is indispensable not only in the folding calculation of
OP and inelastic form factor, but also in the Hartree-Fock (HF) calculations of
nuclear matter \cite{Kh93} and of the finite nuclei \cite{Na03}.

To obtain from the double-folding model presented above the simple expression of
\AA OP used in the OL of Glauber model one needs to make a ``double-zero"
approximation which reduces the complex finite-range $t$-matrix interaction
(\ref{v1}) to a \emph{zero-range} (purely imaginary) NN scattering amplitude at
\emph{zero} NN \emph{angle} $t_{\rm NN}(\theta=0^\circ)\delta(\bm{s})$ that can
be further expressed through the total NN cross section $\sigma_{\rm NN}$, using
the optical theorem. As a result, one needs to evaluate in the OL of Glauber
model only a simple folding integral over local densities of the two colliding
nuclei \cite{To98}
\begin {equation}
 U(\bm{R})\to V_{OL}(\bm{R})=\frac{i\sigma_{\rm NN}}{2}\int\rho_a(\bm{R})
 \rho_A(\bm{R}-\bm{r}_A)d^3r_A. \label{eOL}
\end {equation}

The prescription (\ref{eOL}) is also known as the $t\rho\rho$ approximation
\cite{Hu91} which neglects the off-shell part of the $t$-matrix. Besides the
inaccuracy caused by the use of zero-range approximation \cite{Be89}, the
zero-angle approximation takes into account only the on-shell $t$-matrix at zero
momentum transfer (see Eq.~(3) in Ref.~\cite{Lo81}). Since the
antisymmetrization of $t_{\rm NN}$ requires an accurate estimation of the NN
knock-on exchange term which is strongest at \emph{large} momentum transfers
($q>6$ fm$^{-1}$ at energies around 0.8 GeV \cite{Lo81,Fr85}), the zero-angle
approximation could strongly reduce the strength of the exchange term. A
question remains, therefore, whether the NN antisymmetry is properly taken into
account when one uses the empirical $\sigma_{\rm NN}$ in the Glauber folding
integral (\ref{eOL}). A similar aspect has been raised by Brandan {\sl et al.}
\cite{Mc97} who found that an \emph{overestimated} absorption in the
nucleus-nucleus system (by the $t\rho\rho$ model) is due to the effects of Pauli
principle. To illustrate the importance of the knock-on exchange term, we have
plotted in Fig.~\ref{f0} the \emph{direct} and \emph{exchange} components of the
microscopic OP for $^{6}$He+$^{12}$C system at 790 MeV/nucleon predicted by our
double-folding approach using realistic g.s. densities (see next Sect.) of the
two colliding nuclei. One can see that the exchange term of the real OP is
repulsive and much stronger than the (attractive) direct term, which makes the
total real OP repulsive at all internuclear distances (see panel \emph{a} of
Fig.~\ref{f0}). The exchange term of the imaginary OP is also repulsive but its
relative strength is much weaker compared to that of the real OP, and the total
imaginary OP remains attractive or \emph{absorptive} at all distances. As a
result, the direct part of the imaginary OP is about 10\% more absorptive than
the total imaginary OP (see panel \emph{b} of Fig.~\ref{f0}). The total reaction
cross section predicted by the complex OP shown in Fig.~\ref{f0} is \sigR\
$\approx 727$ mb. This value increases to \sigR\ $\approx 750$ mb when the
exchange potential $V_{\rm EX}$ is omitted in the OM calculation. Consequently,
the relative contribution by the exchange term in \sigR\ is about 3\%. This
difference is not small because it can lead to a difference of up to 7\% in the
extracted nuclear RMS radii. Due to an overwhelming contribution by the exchange
part of the \emph{real} OP, the exchange potential affects the calculated
elastic scattering cross section (see Fig.~\ref{f1}) much more substantially
compared to \sigR, which is determined mainly by the \emph{imaginary} OP.

We will show below a slight (but rather systematic) difference in \sigR\ values
obtained in our approach and the OL of Glauber model that might be due to the
exchange effect. We note further that the elastic $S$-matrix is obtained in our
approach rigorously from the quantal solution of the Schr\"odinger equation for
elastic scattering wave, while the elastic $S$-matrix used in the Glauber model
is given by the eikonal approximation which neglects the second-derivative term
of the same Schr\"odinger equation.

A common feature of the present folding approach and the OL of Glauber model is
the use of single-particle nuclear densities of the projectile and target as
input for the calculation, leaving out all few-body correlations to the
structure model used to construct the density. This simple ansatz has been
referred to as ``static density approximation" \cite{Al96,To98} which does not
take into account explicitly the dynamic few-body correlation between the core
and valence nucleons in a loosely-bound projectile while it collides with the
target. In the Glauber model, this type of few-body correlation can be treated
explicitly \cite{Og92,Al96,To98} using simple assumptions for the wave functions
of the core and valence nucleons as well as that of their relative motion. For
unstable nuclei with a well-extended halo structure, like $^{11}$Li or $^6$He,
such an explicit treatment of the dynamic few-body correlation leads
consistently to a smaller \sigR, i.e., to a larger nuclear radius compared to
that given by the OL of Glauber model \cite{Og92,Al96,To98}. On the level of the
HF-type folding calculation (\ref{e5}), an explicit treatment of the core and
valence nucleons would result in a much more complicated triple-folding
formalism which involves the antisymmetrization not only between the projectile
nucleons and those of the target, but also between the nucleons of the core and
the valence nucleons. Such an approach would clearly end up with a
\emph{nonlocal} OP which will not be easily used with the existing direct
reaction codes. The lack of an appropriate treatment of the dynamic few-body
correlations remains, therefore, the main limitation of the present folding
approach in the calculation of the OP for systems involving unstable nuclei with
halo-type structure.

Note that an effective way of taking into account the loose binding between the
core and valence nucleons is to add a higher-order contribution from breakup
(dynamic polarization potential) to the first-order folded potential
\cite{Kh95b,Kh95a} or simply to renormalize the folded potential to fit the
data. However, validity of the IA implies that higher-order multiple scattering
or contribution from the dynamic polarization potential are negligible, and the
folded OP and inelastic form factor based on the $t$-matrix interaction
(\ref{v1}) should be used in the calculations without any further
renormalization. Therefore, we will discuss below only results obtained with the
unrenormalized folded potentials, keeping in mind possible effects due to the
few-body correlation.

\begin{table*}
\caption{The total reaction cross section \sigR\ and interaction cross section
\sigI\ calculated for stable $^{4}$He, $^{6,7}$Li, $^{12}$C, and $^{16}$O nuclei
in comparison with $\sigma^{\rm exp}_{\rm I}$ taken from the data compilation in
Ref.~\cite{Oz01}.
 $\Delta\sigma_{\rm I}=|\sigma^{\rm calc}_{\rm I}-\sigma^{\rm exp}_{\rm I}|/
 \sigma^{\rm exp}_{\rm I}$.} \label{t2}
\begin{ruledtabular}
\begin{tabular}{|c|c|c|c|c|c|c|c|c|c|} %\hline
Nucleus & Energy & Density Model & \rmscalc & Reference &
 \rmsexp & $\sigma^{\rm calc}_{\rm R}$ & $\sigma^{\rm calc}_{\rm I}$ &
  $\sigma^{\rm exp}_{\rm I}$ & $\Delta\sigma_{\rm I}$  \\
 & (MeV/nucleon) & & (fm) & & (fm) & (mb) & (mb) & (mb) & (\%) \\ \hline
 $^4$He & 790 & HO & 1.461 & \cite{Sa79} & $1.47\pm 0.02\ ^a$ & 513 & 504 &
 $503\pm 5$ & 0.2 \\
  &  & HO & 1.550 & \cite{Kh01} &  & 523 & 515 &  & 2.4 \\
  &  & HO & 1.720 & \cite{So99} &  & 543 & 536 &  & 6.6 \\  \hline
 $^6$Li & 790 & IPM & 2.401 & \cite{Sa80} & $2.43\pm 0.02\ ^a$ & 722 & 717 &
 $688\pm 10$ & 4.2 \\
  &  & HO & 2.401 & This work &  & 723 & 718 &  & 4.4 \\
  &  & HO & 2.320 & This work &  & 709 & 703 &  & 2.2 \\ \hline
 $^7$Li & 790 & IPM & 2.367 & \cite{Sa80} & $2.33\pm 0.02\ ^b$ & 746 & 741 &
 $736\pm 6$  & 0.7 \\
 & & HO & 2.334 & This work &  & 744 & 739 & & 0.4 \\ \hline
 $^{12}$C & 950 & FM & 2.332 & \cite{Kh01} & $2.33\pm 0.02\ ^a$ & 854 & 844 &
 $853\pm 6$ & 1.1 \\
  &  & HO & 2.332 & \cite{Kh01} &  & 853 & 843 &  & 1.1 \\
  &  & HFB & 2.446 & This work &  & 881 & 872 &  & 2.2 \\ \hline
 $^{16}$O & 970 & FM & 2.618 & \cite{Kh01} & $2.61\pm 0.01\ ^a$ & 992 & 981 &
 $982\pm 6$ & 0.1 \\
  &  & HO & 2.612 & \cite{Kh01} &  & 988 & 978 &  & 0.4 \\
  &  & HFB & 2.674 & This work &  & 1006 & 997 &  & 1.4 \\ %\hline
\end{tabular}
\end{ruledtabular}
$^a$ RMS radius of the proton density given by the experimental charge density
\cite{Vr87} unfolded with the finite size of proton. \\ $^b$ Nuclear RMS radius
deduced from the Glauber model analysis of the same \sigI\ data in the OL
approximation \cite{Oz01}.
\end{table*}

\section{Results and discussion}
 \label{sec3}
\subsection*{Results for stable $(N=Z)$ isotopes}
An important step in any experimental or theoretical reaction study with
unstable beams is to gauge the method or model by the results obtained with
stable beams. Therefore, we have considered first the available data of \sigI\
induced by stable $^{4}$He, $^{6}$Li, $^{12}$C, and $^{16}$O beams on $^{12}$C
target \cite{Oz01}. These $(N=Z)$ nuclei are strongly bound, and the RMS radius
of the (point) proton distribution inferred from the elastic electron scattering
data \cite{Vr87} can be adopted as the ``experimental" nuclear radius if the
proton and neutron densities are assumed to be the same. To show the sensitivity
of the calculated \sigI\ to the nuclear radius, we present in Table~\ref{t2}
results obtained with different choices for the projectile density in each case.
We use for the g.s. density of $^{12}$C target a realistic Fermi (FM)
distribution \cite{Kh01}
\begin {equation}
 \rho_0(r)=\rho_0/[1+\exp((r-c)/a)],
\label{s3}
\end {equation}
where $\rho_0=0.194$ fm$^{-3},\ c=2.214$ and $a=0.425$ fm were chosen to
reproduce the shape of shell model density and experimental radius of 2.33 fm
for $^{12}$C.

$^4$He is a unique case where a simple harmonic oscillator (HO) model can
reproduce quite well its ground state density. If one chooses the HO parameter
to give \rmsr = 1.461 fm (close to the experimental radius of $1.47\pm 0.02$ fm)
then one obtains the Gaussian form adopted in Ref.~\cite{Sa79} for
$\alpha$-density. This choice of $^4$He density has been shown in the folding
analysis of elastic \aA scattering \cite{Kh01} to be the most realistic. By
comparing the calculated \sigI\ with the data, we find that this same choice of
$^4$He density gives the best agreement between $\sigma^{\rm calc}_{\rm I}$ and
$\sigma^{\rm exp}_{\rm I}$. Similar situation was found for $^{12}$C and
$^{16}$O isotopes, where the best agreement with the data is given by the
densities which reproduce the experimental nuclear radii. Beside a simple Fermi
distribution \cite{Kh01}, microscopic g.s. densities given by the
Hartree-Fock-Bogoliubov (HFB) calculation that takes into account the continuum
\cite{Gr01} were also used. The agreement with the data for $^{12}$C and
$^{16}$O given by the HFB densities is around 2\%, quite satisfactory for a
fully microscopic structure model. We have further used $sp$-shell HO wave
functions to construct the g.s. densities of $^{6}$Li, $^{12}$C, and $^{16}$O.
For $^{12}$C and $^{16}$O, the best agreement with the \sigI\ data is again
reached when the HO parameter is tuned to reproduce the experimental radii.

The agreement is slightly worse for $^{6}$Li compared to $^{4}$He, $^{12}$C, and
$^{16}$O cases if $^{6}$Li density distribution reproduces the experimental
radius. We have first used $^{6}$Li density given by the independent particle
model (IPM) developed by Satchler \cite{Sa79,Sa80} which generates realistic
wave function for each single-particle orbital using a Woods-Saxon (WS)
potential for the bound state problem. The IPM density gives \rmsr$\approx 2.40$
fm for $^{6}$Li, rather close to the experimental radius of $2.43\pm 0.02$ fm
inferred from $(e,e)$ data \cite{Vr87}. The HO density gives the same \sigI\ as
that given by the IPM density if the HO parameter is chosen to give the same
radius of 2.40 fm. These two versions of $^{6}$Li density overestimate the
\sigI\ data by about 4\%. If the HO parameter is chosen to give \rmsr$\approx
2.32$ fm then the agreement with the \sigI\ data improves to around 2\%. This
result indicates that our folding + DWIA analysis slightly overestimates the
absorption in $^{6}$Li+$^{12}$C system. Since $^{6}$Li is a loosely bound
$\alpha+d$ system, this few percent discrepancy with the \sigI\ data might well
be due to the dynamic correlation between the $\alpha$-core and deuteron cluster
in $^{6}$Li during the collision which is not taken into account by our
approach. Note that a few-body Glauber calculation \cite{To98} (which takes into
account explicitly the dynamic correlation between $\alpha$ and $d$) ends up,
however, with about the same discrepancy (see Fig.~4 in Ref.~\cite{To98}).
$^{6}$Li remains, therefore, an interesting case for the reaction models to
improve their ingredients. For $^{7}$Li, the IPM density \cite{Sa79} gives
\rmsp$\approx 2.28$ fm (close to the experimental value of $2.27\pm 0.01$ fm
\cite{Vr87}) and \rmsn$\approx 2.43$ fm which make the matter radius
\rmsr$\approx 2.37$ fm. As a result, \sigI\ calculated with the IPM density for
$^{7}$Li agrees with the data within less than 1\%. In the HO model for $^{7}$Li
density, we have chosen the HO parameter for protons to reproduce the
experimental radius of 2.27 fm and that for neutrons adjusted by the best
agreement with the \sigI\ data. The best-fit \rmsr radius then becomes around
2.33 fm.

\begin{table*}
\caption{The same as Table~\ref{t2} but for neutron-rich He, Li, C, and O
isotopes. Note that \rmscalc\ given by the HO densities should have about the
same uncertainties as those deduced for \rmsexp\ by the OL of Glauber model.}
\label{t3}
\begin{ruledtabular}
\begin{tabular}{|c|c|c|c|c|c|c|c|c|c|} %\hline
Nucleus & Energy & Density Model & \rmscalc & Reference &
 \rmsexp & $\sigma^{\rm calc}_{\rm R}$ &
 $\sigma^{\rm calc}_{\rm I}$ & $\sigma^{\rm exp}_{\rm I}$ &
 $\Delta\sigma_{\rm I}$  \\
 & (MeV/nucleon) & & (fm) & & (fm) & (mb) & (mb) & (mb) & (\%) \\ \hline
 $^6$He & 790 & HF & 2.220 & \cite{Be89} & $2.48\pm 0.03\ ^a$ & 691 & 686 &
 $722\pm 6$ & 5.0 \\
 & & 3-BODY & 2.530 & \cite{Al96} &  & 738 & 733 &  & 1.5 \\
 & & IPM & 2.460 & This work & $2.45\pm 0.10\ ^b$ & 727 & 722 & & 0.0 \\
 $^8$He & 790 & COSMA & 2.526 & \cite{Zh94} & $2.52\pm 0.03\ ^a$ & 816 & 812 &
 $817\pm 6$ & 0.6 \\ \hline
$^8$Li & 790 & HO & 2.371 & This work & $2.37\pm 0.02\ ^a$ & 782 &
775 &
 $768\pm 9$ & 0.9 \\
 $^9$Li & 790 & HO & 2.374 & This work & $2.32\pm 0.02\ ^a$ & 809 & 802 &
 $796\pm 6$ & 0.7 \\
 $^{11}$Li & 790 & HO+halo & 3.227 & This work & $3.12\pm 0.16\ ^a$ & 1066 &1061 &
 $1060\pm 10\ ^c$ & 0.1 \\
   &   & HF & 2.868 & \cite{Be89} &   & 971 & 967 &  & 8.8 \\ \hline
$^{13}$C & 960 & IPM & 2.389 & \cite{Sa80} &
$2.28\pm 0.04\ ^a$ & 887 & 877 & $862\pm 12$ & 1.7 \\
&  & HO& 2.355 & This work &  & 875 & 866 &  & 0.5 \\
 $^{14}$C & 965 & HFB & 2.585 & This work & $2.30\pm 0.07\ ^a$
 & 951 & 941 & $880\pm 19$ & 6.9 \\
   &   & IPM & 2.417 & \cite{Sa80} &  & 910 & 900 &  & 2.3 \\
       &   & HO & 2.386 & This work &  & 899 & 888 &  & 0.9 \\
$^{15}$C & 740 & HO & 2.481 & This work & $2.40\pm 0.05\ ^a$ & 961
& 952 & $945\pm 10$ & 0.7 \\
 $^{16}$C & 960 & HFB & 2.724 & This work & $2.70\pm 0.03\ ^a$ & 1026 & 1018 &
 $1036\pm 11$ & 1.7 \\
  &   & HO & 2.782 & This work &  & 1039 & 1030 &  & 0.6 \\
$^{17}$C & 965 & HO& 2.831 & This work &
$2.72\pm 0.03\ ^a$ & 1069 & 1060 & $1056\pm 10$ & 0.4 \\
 $^{18}$C & 955 & HFB & 2.860 & This work & $2.82\pm 0.04\ ^a$ & 1102 & 1094 &
 $1104\pm 15$ & 0.9 \\
   &   & HO & 2.900 & This work &  & 1107 & 1098 &  & 0.5 \\
   $^{19}$C&  960& HO & 3.238 & This work & $3.13\pm 0.07\ ^a$ & 1234 & 1227 &
   $1231\pm 28$& 0.3 \\
 $^{20}$C & 905 & HFB & 2.991 & This work & $2.98\pm 0.05\ ^a$ & 1186 & 1179 &
 $1187\pm 20$ & 0.7 \\
   &   & HO & 3.061 & This work &  & 1196 & 1187 &  & 0.0 \\\hline
$^{17}$O & 970 &   IPM & 2.766 & \cite{Sa80} & $2.59\pm 0.05\ ^a$
& 1026 & 1016 & $1010\pm 10$ & 0.6 \\
&   & HO & 2.672 & This work &  & 1021 & 1011 &  & 0.1 \\
 $^{18}$O & 1050 & HFB & 2.763 & This work & $2.61\pm 0.08\ ^a$ & 1053 & 1042 &
 $1032\pm 26$ & 1.0 \\
 &   & IPM & 2.768 & \cite{Sa80} &  & 1057 & 1048 &  & 1.6 \\
 &   & HO & 2.742 & This work &  & 1046 & 1036 &  & 0.4 \\
 $^{19}$O & 970 & HO & 2.774 & This work & $2.68\pm 0.03\ ^a $ & 1076 & 1066 &
 $1066\pm 9$ & 0.0 \\
 $^{20}$O & 950 & HFB & 2.849 & This work & $2.69\pm 0.03\ ^a$ & 1122 & 1112 &
 $1078\pm 10$ & 3.1 \\
 &   & HO & 2.786 & This work &  & 1100 & 1089 &  & 1.0 \\
 $^{21}$O & 980 & HO & 2.811 & This work &$2.71\pm 0.03\ ^a $  & 1116 & 1105 &
 $1098\pm 11$ & 0.6 \\
 $^{22}$O & 965 & HFB & 2.919 & This work & $2.88\pm 0.06\ ^a$ & 1170 & 1159 &
 $1172\pm 22$ & 1.1 \\
 &   & HO & 2.956 & This work &  & 1178 & 1168 &  & 0.3 \\
 $^{23}$O & 960 & HO & 3.286 & This work &$3.20\pm 0.04\ ^a $  & 1310 & 1302 &
 $1308\pm 16$ & 0.5 \\
 $^{24}$O & 965 & HFB & 3.050 & This work & $3.19\pm 0.13\ ^a$ & 1248 & 1238 &
 $1318\pm 52$ & 6.1 \\
  &   & HO & 3.280 & This work &  & 1319 & 1311 &  & 0.5 \\%\hline
\end{tabular}
\end{ruledtabular}
$^a$ Nuclear RMS radius deduced from the Glauber model analysis of the \sigI\
data in the OL approximation
\cite{Oz01}. \\
$^b$ Nuclear RMS radius deduced from the Glauber model analysis of elastic
$^{6}$He+p scattering data at 0.7 GeV/nucleon \cite{Al02}. \\
$^c$ \sigI\ data taken from Ref.~\cite{Ko89}.
\end{table*}

We conclude from these results that the present folding + DWIA approach and
local $t$-matrix interaction by Franey and Love \cite{Fr85} are quite suitable
for the description of the \AA interaction cross section at energies around 1
GeV/nucleon, with the prediction accuracy as fine as 1--2\% for the stable and
strongly bound nuclei.

\subsection*{Results for neutron-rich isotopes}

\begin{figure*}[htb]%\vspace*{-1cm}
\begin{minipage}[t]{8.5cm}
\hspace*{-1cm}
%\vspace*{-4cm}
\mbox{\epsfig{file=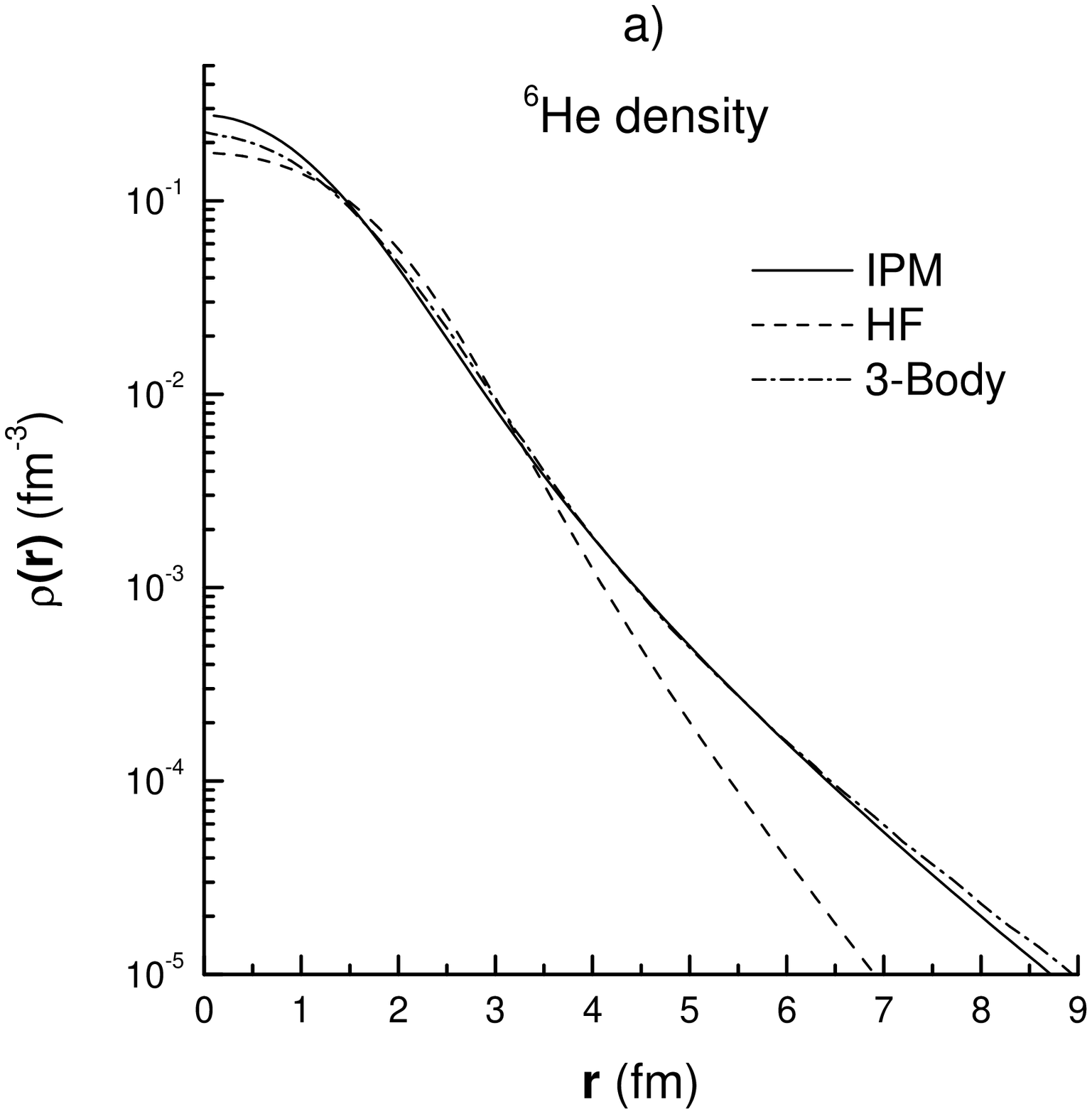,height=13.5cm}}
\end{minipage}
\hspace{\fill}
\begin{minipage}[t]{8.5cm}
\hspace*{-2cm} %\vspace*{-1cm}
 \mbox{\epsfig{file=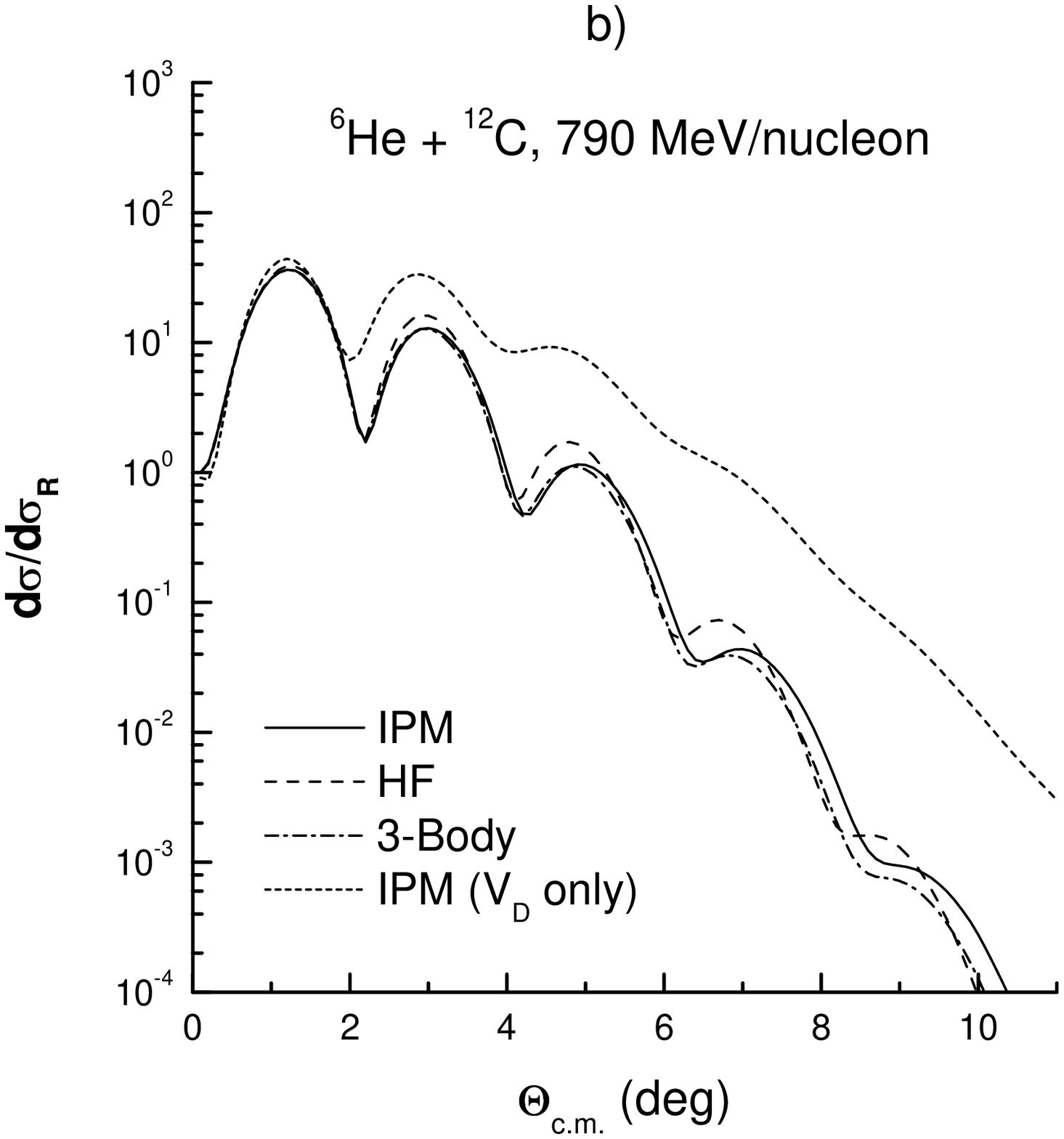,height=13.5cm}}
\end{minipage}
\vspace*{-4cm} \caption{Three versions of $^6$He g.s. density used in the
folding calculation (panel \emph{a}) and elastic $^{6}$He+$^{12}$C scattering
cross sections at 790 MeV/nucleon obtained with the corresponding complex
folded optical potentials (panel \emph{b}). The dotted curve in panel \emph{b}
is obtained without the exchange part of the OP.} \label{f1}
\end{figure*}

Our results for neutron-rich He, Li, C, and O isotopes are presented in
Table~\ref{t3}. Since $^{6}$He beams are now available with quite a good
resolution, this nucleus is among the most studied unstable nuclei. In the
present work we have tested 3 different choices for $^{6}$He density in the
calculation of \sigI. The microscopic $^{6}$He density obtained in a HF
calculation \cite{Be89} has a rather small radius \rmsr$\approx 2.20$ fm and the
calculated \sigI\ underestimates the data by about 5\%. A larger radius of 2.53
fm is given by the density obtained in a consistent three-body formalism
\cite{Al96} and the corresponding \sigI\ agrees better with the data. Given an
accurate $^{7}$Li density obtained in the IPM \cite{Sa79} as shown above and the
fact that $^{6}$He can be produced by a proton-pickup reaction on $^{7}$Li, we
have constructed the g.s. density of $^{6}$He in the IPM (with the recoil effect
properly taken into account \cite{Sa80}) using the following WS parameters for
the single-particle states: $r_0=1.25$ fm, $a=0.65$ fm for the $s\frac{1}{2}$
neutrons and protons which are bound by $S_n=25$ MeV and $S_p=23$ MeV,
respectively; $r_0=1.35$ fm, $a=0.65$ fm for the $p\frac{3}{2}$ halo neutrons
which are bound by $S_n=1.86$ MeV. The WS depth is adjusted in each case to
reproduce the binding energy. The obtained IPM density gives the proton, neutron
and total nuclear radii of $^{6}$He as 1.755, 2.746 and 2.460 fm, respectively.
This choice of $^{6}$He density also gives the best agreement with the \sigI\
data. We note that a Glauber model analysis of the elastic $^{6}$He+p scattering
at 0.7 GeV/nucleon \cite{Al02}, which takes into account higher-order multiple
scattering effects, gives a best-fit \rmsr$\approx 2.45$ fm for $^{6}$He, very
close to our result. Since elastic $^{6}$He+$^{12}$C scattering has recently
been measured at lower energies \cite{La02}, we found it interesting to plot the
3 densities and elastic $^{6}$He+$^{12}$C scattering cross sections at 790
MeV/nucleon predicted by the corresponding complex folded OP (the radial shape
of the OP obtained with the IPM density for $^{6}$He is shown in Fig.~\ref{f0}).
As can be seen from Fig.~\ref{f1}, the IPM density has the neutron-halo tail
very close to that of the density calculated in the three-body model \cite{Al96}
and they both give a good description of \sigI. The predicted elastic cross
section is strongly forward peaked and the difference in densities begins to
show up after the first diffractive maximum. Such a measurement should be
feasible at the facilities used for elastic $^{6}$He+p scattering at 0.7
GeV/nucleon \cite{Al02} and would be very helpful in testing finer details of
$^{6}$He density. As already discussed in previous Sect., the exchange part of
the microscopic OP affects the elastic cross section very strongly (see dotted
curve in panel \emph{b} of Fig.~\ref{f1}) and the elastic $^{6}$He+$^{12}$C
scattering measurement would be also a very suitable probe of the exchange
effects in this system.

Since $^{6}$He is a loosely bound halo nucleus with a well established
three-body $\alpha+n+n$ structure, the dynamic correlation between the
$\alpha$-core and dineutron is expected to be important during the collision.
Our folding + DWIA approach using three-body density for $^{6}$He (version FC
\cite{Al96}) gives \sigI\ $\approx 733$ mb compared to about 720 mb given by the
few-body calculation by Tostevin {\sl et al.} (see Fig.~4 in Ref.~\cite{To98})
based on the same three-body wave function for $^{6}$He. The difference in the
calculated \sigI\ leads to an increase of about 2-3\% in the \rmsr value. It is
likely that such a difference is, in part, due to the dynamic correlation
between the $\alpha$-core and dineutron which was not considered in our folding
+ DWIA approach. For $^{8}$He nucleus, the OL of Glauber analysis of \sigI\ data
\cite{Oz01} and the multiple-scattering Glauber analysis of elastic $^{8}$He+p
data at 0.7 GeV/nucleon \cite{Al02} give \rmsr around 2.52 and 2.53 fm,
respectively. By using the microscopic $^{8}$He density obtained in a four-body
(COSMA) model \cite{Zh94}, which gives \rmsr = 2.526 fm, our folding + DWIA
approach reproduces the measured \sigI\ data within less than 1\%. Note that a
(multiple scattering) Glauber model analysis of the elastic $^{6,8}$He+p
scattering at 0.7 GeV/nucleon which takes into account the dynamic few-body
correlation explicitly was done by Al-Khalili and Tostevin \cite{Al98}, and they
have obtained the best-fit nuclear radii of about 2.5 and  2.6 fm for $^{6}$He
and $^{8}$He, respectively, around 2\% larger than our results.

\begin{table*}
\caption{The HO-density parameters (\ref{r2}) for neutron-rich Li, C, and O
isotopes.} \label{t4}
\begin{ruledtabular}
\begin{tabular}{|c|c|c|c|c|c|c|c|c|c|} %\hline
Nucleus & $P_n$ & $P_p$ & $D_n$ & $D_p$ & $b_n$ & $b_p$ & \rmsn &
 \rmsp & \rmsr \\
 & & & & & (fm) & (fm) & (fm) & (fm) & (fm) \\ \hline
 $^7$Li  & 2/3& 1/3 &0.0 &0.0 & 1.684 & 1.6766 & 2.382 & 2.270 & 2.334 \\
 $^8$Li  &1.0 &1/3 &0.0 &0.0 & 1.6770 & 1.6776 & 2.430 &2.270 & 2.371 \\
 $^9$Li  &4/3 &1/3 &0.0 &0.0 & 1.6470 & 1.6766 & 2.424 & 2.270 & 2.374 \\ \hline
$^{13}$C  &5/3 &4/3 &0.0 &0.0 & 1.6058 &1.5722 & 2.389 & 2.314 & 2.355\\
 $^{14}$C & 2.0&4/3 &0.0 &0.0 & 1.6226 & 1.5762 & 2.434 & 2.320 & 2.386 \\
 $^{15}$C & 2.0&4/3 &2/15 &0.0 & 1.6630 & 1.5898 & 2.570 & 2.340 & 2.481 \\
 $^{16}$C & 2.0&4/3 &4/15&0.0 &1.8512 & 1.7128 & 2.927 & 2.521 & 2.782 \\
 $^{17}$C  & 2.0&4/3 &2/5 &0.0 & 1.8552& 1.7128 & 2.986 & 2.521 & 2.831 \\
 $^{18}$C & 2.0&4/3 &8/15 &0.0 & 1.8752 & 1.7297 & 3.062 & 2.546 & 2.900 \\
 $^{19}$C & 2.0&4/3 &2/3  &0.0 & 2.1252 & 1.7297 & 3.512 & 2.546 & 3.238 \\
 $^{20}$C & 2.0&4/3 &4/5  &0.0 & 1.9462 & 1.7467 & 3.248 & 2.571 & 3.061 \\ \hline
 $^{17}$O  &2.0 &2.0 &2/15 & 0.0& 1.7775 & 1.7232 & 2.747 & 2.585 & 2.672 \\
 $^{18}$O &2.0 &2.0 &4/15 & 0.0 & 1.7601 & 1.7935 & 2.783 & 2.690 & 2.742 \\
 $^{19}$O &2.0 &2.0 &2/5  & 0.0 & 1.7601 & 1.7935 & 2.833 & 2.690 & 2.774 \\
 $^{20}$O &2.0 &2.0 &8/15 & 0.0 & 1.7401 & 1.8005 & 2.842 & 2.701 & 2.786 \\
 $^{21}$O &2.0 &2.0 &2/3  & 0.0 & 1.7401 & 1.8005 & 2.876 & 2.701 & 2.811 \\
 $^{22}$O &2.0 &2.0 &4/5  & 0.0 & 1.8498 & 1.8081 & 3.087 & 2.712 & 2.956 \\
 $^{23}$O &2.0 &2.0 &14/15 & 0.0 & 2.1118 & 1.8081 & 3.555 & 2.712 & 3.286 \\
$^{24}$O  &2.0 &2.0 &16/15 & 0.0 & 2.0758 & 1.8261 & 3.520 & 2.739& 3.280 \\
\end{tabular}
\end{ruledtabular}
\end{table*}

\subsubsection*{Parameters of HO densities deduced from \sigI\ data}

Although the HO model is a very simple approach, the HO densities were shown
above to be useful in testing the nuclear radii for stable ($N=Z$) nuclei.
Moreover, the HO-type densities (with the appropriately chosen HO lengths) for
the $sd$-shell nuclei have been successfully used in the analysis of (e,e) data,
measurements of isotope shift and muonic atoms \cite{Oz01}. Therefore, it is not
unreasonable to use simple HO parameterization for the g.s. densities of
neutron-rich nuclei to estimate the nuclear radii, based on our folding + DWIA
analysis of \sigI\ data. For a $N\neq Z$ nucleus, one needs to generate proton
and neutron densities separately as
\begin {equation}
 \rho_\tau(r)=\frac{2}{\pi^{3/2}b^3_\tau}
 \left(1+P_\tau\frac{r^2}{b^2_\tau}+D_\tau\frac{r^4}{b^4_\tau}\right)
 \exp\left(-\frac{r^2}{b^2_\tau}\right),
 \label{r2}
\end{equation}
where $\tau=n$ or $p$, parameters $P_\tau$ and $D_\tau$ are determined from the
nucleon occupation of the $p$- and $d$ harmonic-oscillator shells, respectively.

\begin{figure}[htb]
\hspace*{-1cm} \vspace{-1cm} \mbox{\epsfig{file=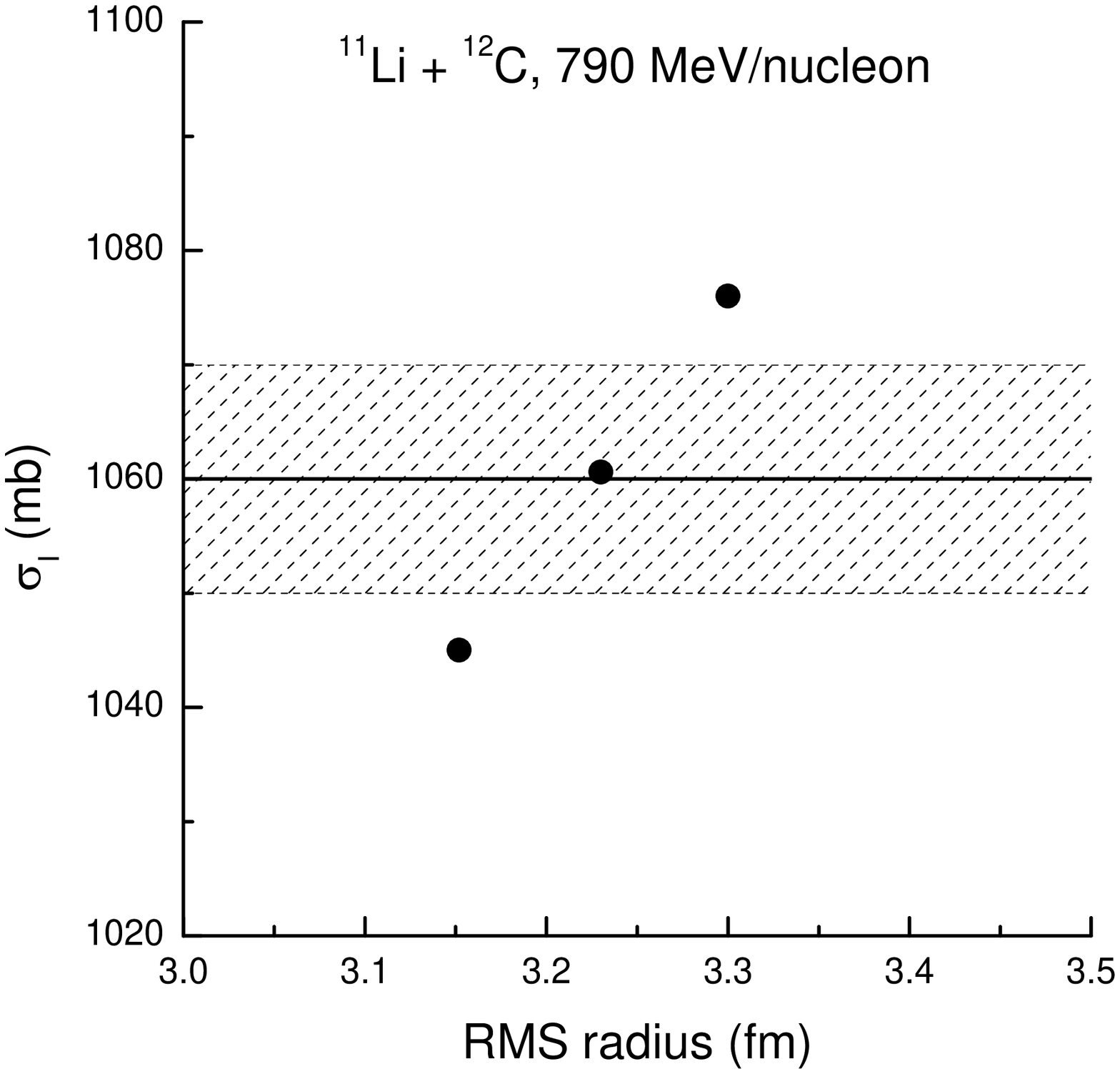,height=13.5cm}}
\vspace{-2cm} \caption{$\sigma^{\rm calc}_{\rm I}$ obtained with three versions
of $^{11}$Li g.s. density, where Gaussian range of the 2$n$-halo was adjusted to
give \rmsr=3.15, 3.23 and 3.30 fm for $^{11}$Li, in comparison with
 $\sigma^{\rm exp}_{\rm I}=1060\pm 10$ mb \cite{Ko89}.} \label{f2}
\end{figure}

To generate the g.s. densities of $^{8,9}$Li isotopes, we have assumed the
proton density of these nuclei to be approximately that of $^7$Li and the
neutron HO length $b_n$ is adjusted in each case to reproduce the measured
\sigI\ (see Tables~\ref{t3} and \ref{t4}). While the obtained \rmsr for $^{8}$Li
is rather close to that given by the OL of Glauber model \cite{Oz01}, results
obtained for $^9$Li are different and we could reproduce the \sigI\ data only if
the neutron HO length is chosen to give \rmscalc $\approx 2.37$ fm or about 2\%
larger than that given by the OL of Glauber model. For the halo nucleus
$^{11}$Li, a $^9$Li core + two-neutron halo model was used to generate its
density. Namely, we have used HO density of $^9$Li that reproduces the measured
\sigI\ for $^9$Li and a Gaussian tail for the two-neutron halo density. To reach
a best agreement between $\sigma^{\rm exp}_{\rm I}$ taken from Ref.~\cite{Ko89}
and $\sigma^{\rm calc}_{\rm I}$, the Gaussian range was chosen to give \rmscalc
$\approx 3.23$ fm which is about 0.1 fm larger than that given by the OL of
Glauber model \cite{Oz01}. A microscopic density for $^{11}$Li obtained in the
HF calculation \cite{Be89} (which gives \rmsr = 2.868 fm) has also been used in
our folding analysis. The agreement with the data becomes much worse in this
case (see Table~\ref{t3}) and we conclude that the radius given by the HF
density is somewhat too small. To show the sensitivity of our analysis to the
nuclear radius, we have plotted in Fig.~\ref{f2} \sigI\ predicted by three
versions of $^{11}$Li density with the Gaussian range of the $2n$-halo adjusted
to give \rmsr = 3.15, 3.23 and 3.30 fm, respectively, compared to $\sigma^{\rm
exp}_{\rm I}=1060\pm 10$ mb \cite{Ko89}. It is easily to infer from
Fig.~\ref{f2} an empirical RMS radius of $3.23\pm 0.05$ fm for $^{11}$Li. Note
that \sigI\ measurement for $^{11}$Li+$^{12}$C system at 790 MeV/nucleon has
been reported in several works with $\sigma^{\rm exp}_{\rm I}=1040\pm 60$
\cite{Ta85}, $1047\pm 40$ \cite{Ta88} and $1060\pm 10$ mb \cite{Ko89}. If we
adjust Gaussian range of the $2n$-halo in $^{11}$Li density to reproduce these
$\sigma^{\rm exp}_{\rm I}$ values, the corresponding \rmsr radii of $^{11}$Li
are 3.13, 3.15 and 3.23 fm, respectively. Since \sigI\ data obtained in
Ref.~\cite{Ko89} has a much better statistics and less uncertainty, we have
adopted \rmsr = $3.23\pm 0.05$ fm as the most realistic RMS radius of $^{11}$Li
given by our folding + DWIA analysis.

The total reaction cross section for $^{11}$Li+$^{12}$C system at 790
MeV/nucleon has been studied earlier in the few-body Glauber formalism by
Al-Khalili {\sl et al.} \cite{Al96}, where \rmsr radius for $^{11}$Li was shown
to increase from 3.05 fm (in the OL) to around 3.53 fm when the dynamic
correlation between $^{9}$Li-core and $2n$-halo during the collision is treated
explicitly. This is about 9\% larger than \rmsr radius obtained in our folding +
DWIA approach based on the same \sigI\ data. Although various structure
calculations for $^{11}$Li give its RMS radius around 3.1--3.2 fm (see
Refs.~\cite{Oz01,Ta96} and references therein), a very recent coupled-channel
three-body model for $^{11}$Li by Ikeda {\em et al.} \cite{Ik03,My03} shows that
its RMS radius is ranging from 3.33 to 3.85 fm if the $2n$-halo wave function
consists of 21 to 39\% mixture from $(s\frac{1}{2})^2$ state, respectively. A
comparison of the calculated Coulomb breakup cross section with the data
\cite{My03} suggests that this $s$-wave mixture is around 20--30 \%. Thus, the
nuclear radius of $^{11}$Li must be larger than that accepted sofar
\cite{Oz01,Ta96} and be around 3.3--3.5 fm, closer to the result of the few-body
calculation \cite{Al96} and the upper limit of RMS radius given by our folding +
DWIA analysis.

\begin{figure*}[htb]%\vspace*{-1cm}
\begin{minipage}[t]{8.5cm}
\hspace*{-1cm}
%\vspace*{-4cm}
\mbox{\epsfig{file=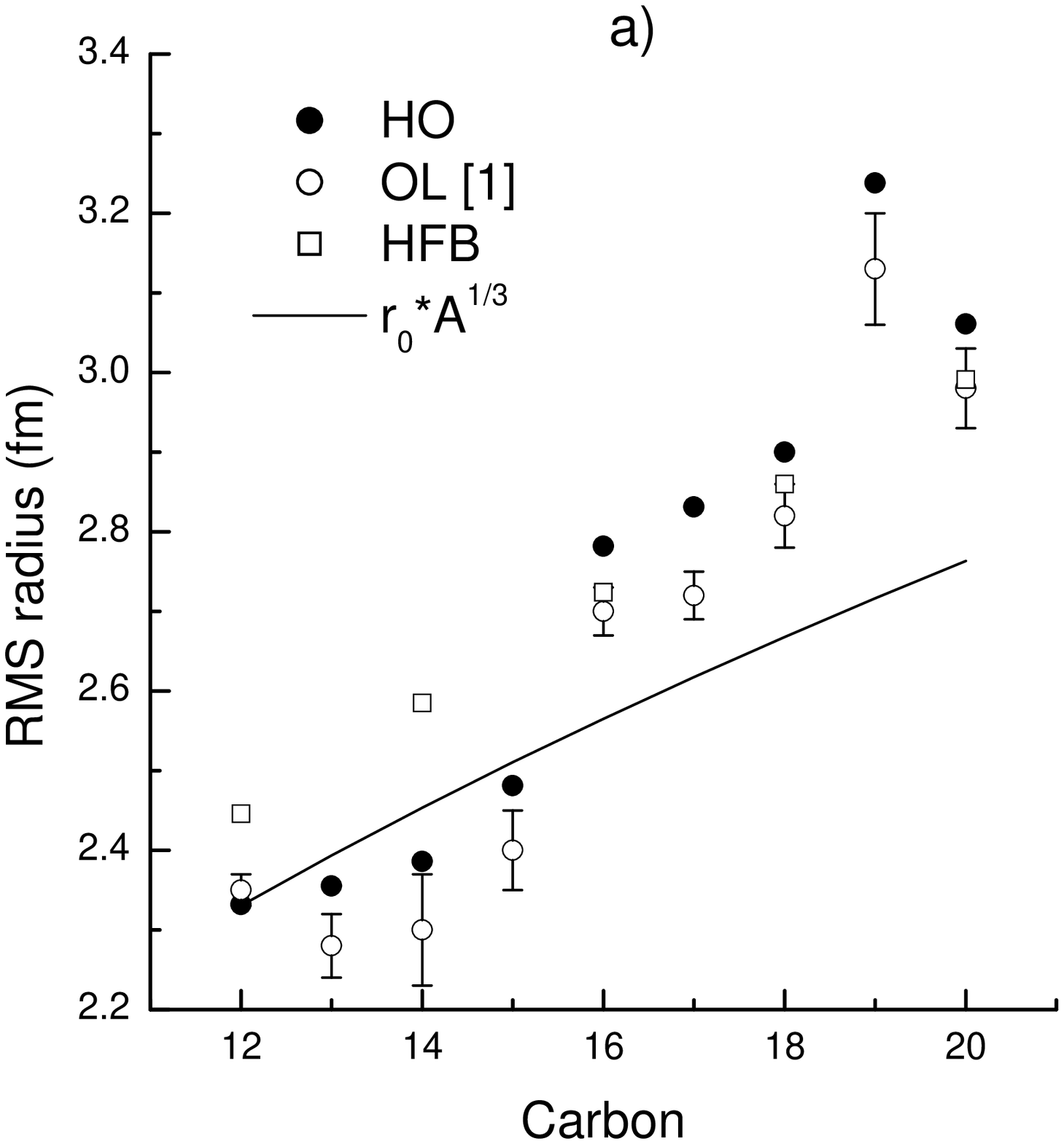,height=13.5cm}}
\end{minipage}
\hspace{\fill}
\begin{minipage}[t]{8.5cm}
\hspace*{-1cm} %\vspace*{-1cm}
 \mbox{\epsfig{file=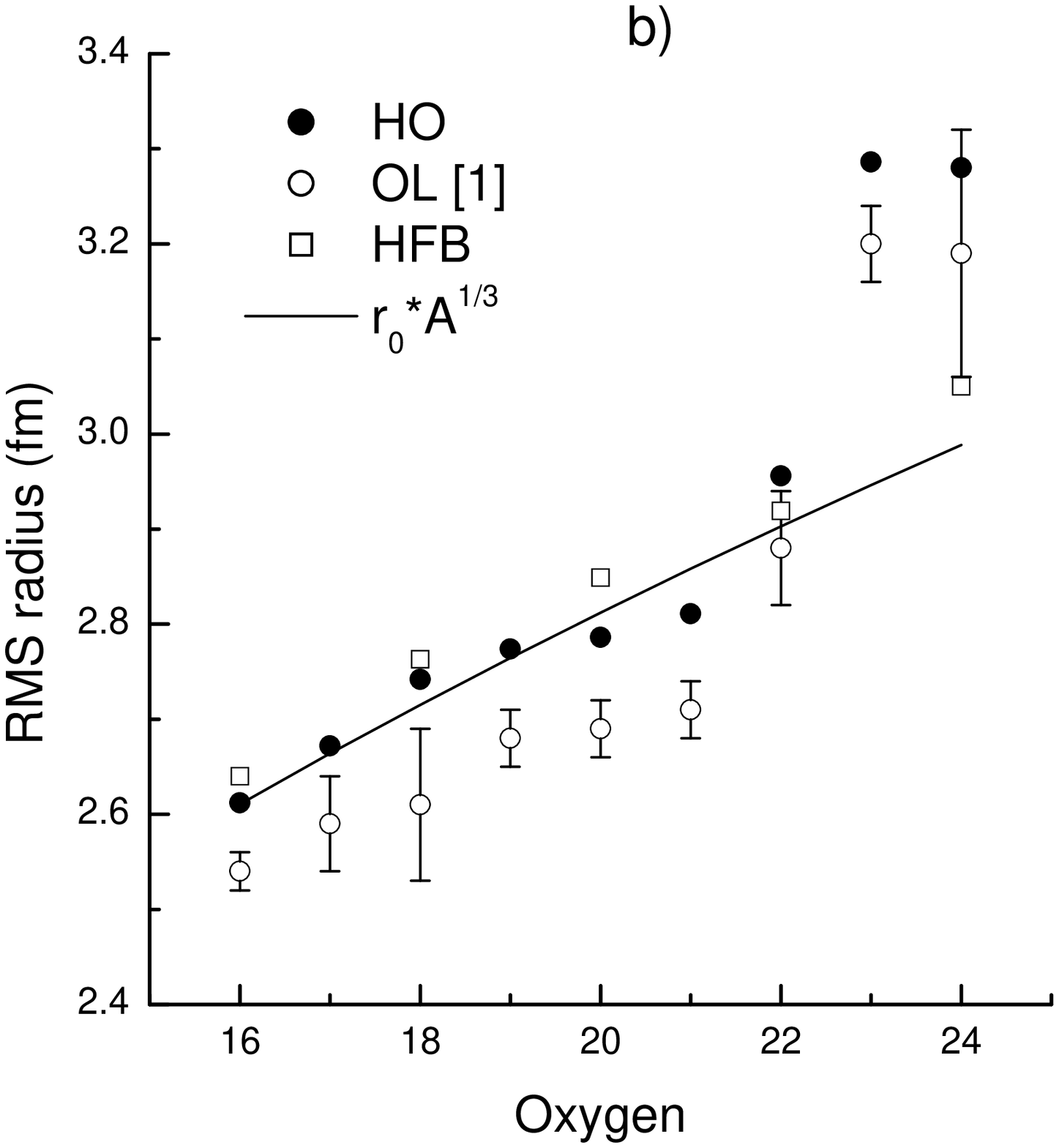,height=13.5cm}}
\end{minipage}
\vspace*{-4cm} \caption{Mass dependence of the nuclear RMS radius for Carbon
(panel \emph{a}) and Oxygen (panel \emph{b}) isotopes given by the two choices
(HFB and HO) of the g.s. densities compared to that deduced from the Glauber
model analysis in the OL approximation \cite{Oz01}. The lines represent
$r_0A^{1/3}$ dependence with $r_0$ deduced from the experimental radii of
$^{12}$C and $^{16}$O given in Table~\ref{t2}.} \label{f3}
\end{figure*}

For most of neutron-rich C and O isotopes considered here, we have first fixed
the proton HO lengths $b_p$ to reproduce the proton \rmsp radii predicted by
the microscopic IPM and HFB densities (as described below). The neutron HO
lengths $b_n$ are then adjusted to the best agreement with \sigI\ data, and the
obtained HO parameters are summarized in Table~\ref{t4}.

\subsubsection*{Microscopic HFB densities}

Before discussing the results obtained for the neutron-rich C and O isotopes, we
give here a brief description of the microscopic HFB approach used to calculate
the g.s. densities of even C and O isotopes. More details about this approach
can be found in Ref.~\cite{Gr01}.

We solve the HFB equations in coordinate representation and in spherical
symmetry with the inclusion of continuum states for neutron-rich nuclei. As the
neutron Fermi energies of these nuclei are typically quite close to zero,
pairing correlations can easily scatter pairs of neutrons from the bound states
towards continuum states. For this reason, the inclusion and the treatment of
continuum states in the calculation are very important. In our calculation the
continuum is treated exactly, i.e., with the correct boundary conditions for
continuum wave functions and by taking into account the widths of the
resonances. Resonant states are localized by studying the behavior of the phase
shifts with respect to the quasi-particle energy for each partial wave $(l,j)$.

The calculations were done with the Skyrme interaction SLy4 for the mean field
channel and with the following zero-range density-dependent interaction
\begin {equation}
 V=V_0 \left[ 1- \left(\frac{\rho(r)}{\rho_0} \right)^{\gamma}
 \right] \delta(\bm{r}_1-\bm{r}_2)
\label{*}
\end{equation}
for the pairing channel. In Eq.~(\ref{*}), $\rho_0$ is the saturation density
and $\gamma$ is chosen equal to 1. We have adapted the prescription of
Refs.~\cite{1*,2*} to finite nuclei in order to fix $V_0$ together with the
quasi-particle energy cutoff. This prescription, requiring that the free
neutron-neutron scattering length has to be reproduced in the truncated space,
allows us to deduce a relation between the parameter $V_0$ and the quasiparticle
energy cutoff.

\subsubsection*{Nuclear radii of Carbon and Oxygen isotopes}

The \sigI\ data for neutron-rich C and O isotopes are compared in Table~\ref{t3}
with \sigI\ predicted by different choices of nuclear densities. We have tested
first the IPM density for $^{13}$C \cite{Sa80} based on the single-particle
spectroscopic factors obtained in the shell model by Cohen and Kurath
\cite{Co67}. This IPM density gives \rmsr $\approx 2.39$ fm for $^{13}$C and the
predicted \sigI\ agrees with the data within less than 2\%. We have further made
IPM calculation for $^{14}$C based on the same single-particle configurations,
with the WS parameters for $sp$-shells appropriately corrected for the recoil
effects and experimental nucleon separation energies $S_{n,p}$ of $^{14}$C. This
IPM density gives \rmsr $\approx 2.42$ fm for $^{14}$C and the predicted \sigI\
also agrees with the data within 2\%. The HO densities were also parameterized
for $^{13,14}$C with the proton HO lengths $b_p$ chosen to reproduce \rmsp
values predicted by the IPM. The best-fit neutron HO lengths $b_n$ result in
\rmsr = 2.36 and 2.39 fm for $^{13}$C and $^{14}$C, respectively. These values
agree fairly with those given by the IPM densities. The microscopic HFB density
gives for $^{14}$C a significantly larger \rmsr radius of 2.59 fm and the
calculated \sigI\ overestimates the data by nearly 7\%. Note that the OL of
Glauber model gives smaller radius of 2.28 and 2.30 fm for $^{13}$C and
$^{14}$C, respectively, based on the same \sigI\ data \cite{Oz01}. This means
that the absorption given by the OL of Glauber model is indeed stronger than
that given by our approach, as expected from discussion in Sect.~\ref{sec2}.

For the neutron-rich even $^{16-20}$C isotopes, the HFB densities give a
remarkably better agreement with the data and it is, therefore, reasonable to
fix the proton HO lengths of the HO densities for each of $^{15-20}$C isotopes
to reproduce \rmsp radius predicted by the HFB calculation for the nearest even
neighbor. The best-fit neutron HO lengths result in the nuclear radii quite
close to those given by the HFB densities (see Tables~\ref{t3} and \ref{t4}). We
emphasize that the nuclear radii given by our analysis, using the HO densities
for C isotopes, are about 0.1 fm larger than those deduced from the OL of
Glauber model \cite{Oz01}. Given a high sensitivity of \sigI\ data to the
nuclear size, a difference of 0.1 fm is not negligible.

To illustrate the mass dependence of the nuclear radius, we have plotted in
Fig.~\ref{f3}a the RMS radii given by the two sets (HFB and HO) of the g.s.
densities for C isotopes together with those deduced from the OL of Glauber
model based on the same \sigI\ data \cite{Oz01}. One can see that our result
follows closely the trend established by the OL of Glauber model, although the
absolute \rmsr radii obtained with the HO densities are in most cases larger
than those deduced from the OL of Glauber model. With the exception of the
$^{14}$C case, the radii of even C isotopes given by the microscopic HFB
densities agree reasonably well with the empirical HO results. We have also
plotted in Fig.~\ref{f3} the lines representing $r_0A^{1/3}$ dependence with
$r_0$ deduced from the experimental radii of $^{12}$C and $^{16}$O given in
Table~\ref{t2}. One can see that the behavior of nuclear radius in C isotopes is
quite different from the $r_0A^{1/3}$ law. While \rmsr radii found for
$^{12-15}$C agree fairly with the $r_0A^{1/3}$ law, those obtained for
$^{16-20}$C are significantly higher. In particular, a jump in the \rmsr value
was found in $^{16}$C compared to those found for $^{12-15}$C. This result seems
to support the existence of a neutron halo in $^{16}$C as suggested from the
\sigR\ measurement for this isotope at 85 MeV/nucleon \cite{Zh02}. We have
further obtained a nuclear radius of 3.24 fm for $^{19}$C which is significantly
larger than that found for $^{20}$C. This result might also indicate to a
neutron halo in this odd C isotope.

Situation is a bit different for O isotopes, where the best-fit \rmsr radii
follow roughly the $r_0A^{1/3}$ law up to $^{22}$O. For the stable $^{17,18}$O
isotopes, the IPM densities \cite{Sa80} provide a very good description of the
\sigI\ data (within 1--2\%). The best-fit HO densities give \rmsr radii of 2.67
and 2.74 fm for $^{17}$O and $^{18}$O, respectively, which are rather close to
those given by the IPM densities. Predictions given by the microscopic HFB
densities are also in a good agreement with the data for even O isotopes
excepting the $^{24}$O case, where the HFB density gives obviously a too small
\rmsr radius. Since the HFB calculation already takes into account the continuum
effects \cite{Gr01}, such a deficiency might be due to the static deformation of
$^{24}$O. A jump in the \rmsr value was found for $^{23}$O which could indicate
to a neutron halo in this isotope. Behavior of \rmsr radii given by the best-fit
HO densities agrees with the trend established by the OL of Glauber model
\cite{Oz01} but, like the case of C isotopes, they are about 0.1 fm larger than
those deduced from the OL of Glauber model. Thus, the OL of Glauber model seems
to consistently overestimate \sigR\ for the neutron-rich C and O isotopes under
study in comparison with our approach.

One clear reason for the difference between our results and those given by the
OL of Glauber model analysis is that one has matched directly the calculated
\sigR\ with the measured \sigI\ in the Glauber model analysis \cite{Oz01} to
deduce the nuclear radius. If we proceed the same way with the HO densities for
the considered nuclei, the best-fit \rmsr radii decrease slightly but are still
larger than those given by the OL of Glauber model. As already discussed in
Sect.~\ref{sec2}, the \emph{zero-angle} approximation for the NN scattering
amplitude used in the Glauber model might reduce significantly the strength of
the exchange part of the imaginary OP given by Eq.~(\ref{eOL}) and could
overestimate, therefore, the absorption in the dinuclear system. This effect
should be much stronger if one uses a realistic \emph{finite-range}
representation of the NN scattering amplitude. Bertsch {\sl et al.} have shown
\cite{Be89} that the \emph{zero-range} approximation for the NN scattering
amplitude leads to a reduction of the calculated \sigR\ or an enhancement of the
nuclear radius by a few percent (see Figs.~2 and 3 in Ref.~\cite{Be89}). Thanks
to such a cancellation of the exchange effects by the zero-range approximation
for NN scattering amplitude, the simple OL of Glauber model was able to deliver
reasonable estimates of nuclear radii for many stable and unstable isotopes
\cite{Oz01}. It should be noted that the eikonal approximation for the
scattering wave function used in the Glauber model was introduced in the past to
avoid large numerical calculations. With the computing power available today,
there is no problem to perform the OM and DWIA calculations for different \AA
systems involving large numbers of partial waves, and the folding + DWIA method
presented here can be recommended as a reliable microscopic approach to predict
the elastic scattering cross section and to deduce the nuclear radius from the
measured \sigI.

\section{Conclusion}
 \label{sec4}
In this work we have explored the reliability of the optical model + DWIA
approach as a tool for extracting important information on nuclear sizes from
interaction cross section measurements. We concentrate on the energy region of
0.8 to 1 GeV/nucleon where interaction cross section data exist for various
combinations of stable as well as unstable projectiles on different targets. At
these bombarding energies our knowledge of the empirical optical potential is
scarce, especially for unstable systems, and we have used, therefore, the
folding model to calculate the microscopic (complex) optical potential and
inelastic form factors necessary for our analysis.

We have chosen for the folding input the fully finite-range $t$-matrix
interaction developed by Franey and Love \cite{Fr85}. The folded optical
potentials and inelastic form factors are used further as inputs for the
standard optical model and DWIA calculations of total reaction cross sections
and interaction cross sections induced by stable and unstable He, Li, C, and O
isotopes on $^{12}$C target. By using the well tested nuclear g.s. densities for
the stable $^4$He, $^{12}$C and $^{16}$O isotopes, we found that the Franey and
Love $t$-matrix gives extremely good account of the measured \sigI\ for these
nuclei.

We have further used the nuclear g.s. densities obtained in various structure
models to calculate \sigI\ and made realistic estimate for the nuclear radii of
(still poorly known) neutron-rich isotopes based on the comparison between
$\sigma^{\rm calc}_{\rm I}$ and $\sigma^{\rm exp}_{\rm I}$. For the chains of C
and O isotopes, our results agree reasonably well with the empirical trend
established by the OL of Glauber model \cite{Oz01}, but give consistently larger
\rmsr radii for these nuclei. Such an effect could be due to the unsatisfactory
treatment of the exchange part of the nucleus-nucleus OP in the Glauber model
calculation.

Although the nuclear radii deduced by our approach for some light halo nuclei
might be a few percent smaller than realistic values because the dynamic
few-body correlation was not considered explicitly in the present folding + DWIA
formalism, this fully microscopic approach was shown to be more accurate than
the OL of Glauber model. Given realistic nuclear densities, it can give a
reliable (parameter-free) prediction of the nucleus-nucleus optical potential at
energies around 1 GeV/nucleon. Therefore, it provides the necessary link to
relate the calculated \sigI\ to the nuclear density and RMS radius.

\begin{acknowledgments}
The authors thank G.R. Satchler for making the DOLFIN code available to them
for the IPM calculation of nuclear g.s. densities, W. von Oertzen for helpful
discussion, and W.G. Love for critical remarks to the manuscript. We also thank
A. Ozawa, H. Sagawa, and I.J. Thompson for their correspondence on the nuclear
densities. The research has been supported, in part, by the Natural Science
Council of Vietnam.
\end{acknowledgments}

\end{document}